\pdfoutput=1
\documentclass[twocolumn,showpacs,preprintnumbers,amsmath,amssymb]{revtex4}

\usepackage{xspace}
\usepackage{epsfig}
\usepackage{amsfonts}
\usepackage{graphicx}
\usepackage{url}

\usepackage{subfigure}
\usepackage[dvipdfm,%
              colorlinks,%
          linkcolor=blue,%
        citecolor = blue,%
              hyperindex,%
        plainpages=false,%
         urlcolor = blue,%
       pdfstartview=FitH]{hyperref}

\renewcommand{\~}{$\sim$}

\begin{document}

\title{\textsf{ High field transport in graphene }}

\author{Tian Fang}
\author{Aniruddha Konar}
\author{Huili Xing}
\author{Debdeep Jena}
\email[Electronic mail: ]{djena@nd.edu}

\affiliation{Department of Electrical Engineering and Physics, University of Notre Dame, IN, 46556, USA}

\date{\today}

\begin{abstract}
Transport of carriers in two dimensional graphene at high electric fields is investigated by combining semi-analytical and Monte-Carlo methods.  A semi-analytical high-field transport model based on the high rate of optical phonon emission provides useful estimates of the saturation currents in graphene.  For developing a more accurate picture, the non-equilibrium (hot) phonon effect and the role of electron-electron scattering were studied using Monte Carlo simulations.  Monte Carlo simulations indicate that the hot phonon effect plays a dominant role in current saturation, and electron-electron scattering strongly thermalizes the hot carrier population in graphene.  We also find that electron-electron scattering removes negative differential resistance in graphene.  Transient phenomenon as such as velocity overshoot can be used to speed up graphene-based high speed electronic devices by shrinking the channel length below 80nm if electrostatic control can be exercised in the absence of a band gap.
\end{abstract}



\maketitle
\section{Introduction}
Monolayer graphene has attracted great amount of interest due to its novel electronic properties \cite{geim07natmat, novo05nat, science06gatech, prl07kim}.  It has perfect two-dimensional (2D) geometry and a linear band structure near Dirac points.  Since its discovery, a substantial body of work has been done on charge transport properties in graphene \cite{prB07_gr_impurity,prl08_gr_mobility,prB08_gr_acousticphonon,prl08_suspendedgraphene,condmat09_SO_Aniruddha}.  Electron mobility as high as $\sim$ 120,000 cm$^{2}$/V$\cdot$s at $\sim$ 240 K has been measured in suspended graphene \cite{prl08_suspendedgraphene}.  At high field bias, epitaxial graphene field effect transistors (FETs) on SiC substrates have exhibited high current drives, exceeding \~ 3 mA/$\mu$m \cite{EDL09grfet_HRL}. The current carrying capability of graphene, and the limiting scattering mechanisms have been the focus of recent research.

The earliest studies of high-field current drive suggested that monolayer graphene should be able to deliver currents in the range of \~ 4 mA/$\mu$m by considering the intrinsic optical phonon scattering \cite{DRC08DJ}.  Shortly thereafter, current saturation was experimentally achieved in FET structures and the saturation current was observed to be proportional to the square root of the channel carrier density \cite{08natnano}.  Surface-optical (SO) phonon scattering rather than intrinsic optical phonon scattering was suggested as the dominant scattering mechanism for current saturation in graphene on SiO$_{2}$ substrates.  The effect of hot phonons on limiting the current drive in graphene was then considered, but found to be masked by other scattering mechanisms for graphene on SiO$_{2}$ substrate \cite{prl_highfield_spain}.  Subsequent theoretical and experimental work further confirmed that the limiting mechanism for high-field transport in graphene on SiO$_{2}$ substrates is SO phonon scattering \cite{apl09_highfield_graphene,prb10_SO_IBM,prl_highfieldSOphonon_PSU,DRC2010_Pop}.

The extrinsic scattering from SiO$_{2}$ could in principle be diminished by using different substrates, or by suspending graphene. The saturation current in {\em intrinsic} graphene is still poorly understood.  More recently, optical phonon temperatures as high as 1000 K was reported by Raman measurements in graphene sheets under high bias \cite{nanolett2009_heattrans_grFET}, and the hot phonon effect was observed in experiments \cite{nanolett2010_hotphonon_graphene}.  These findings show the necessity of considering hot phonon effect on the intrinsic high field transport properties of graphene.  Moreover, although electron-electron (e-e) scattering degrades the low field transport properties of graphene \cite{apl10_eescattering_graphene}, its effect on high field transport has not been studied yet.  As a result, the high-field current carrying capacity of intrinsic graphene, considering the hot phonon effect and e-e scattering, is not clearly understood.

In this work, high field carrier transport in intrinsic graphene is investigated by Monte-Carlo (MC) simulations, by including hot phonon effects as well as e-e scattering.  Furthermore, transient high-field transport effects, such as velocity overshoot and transit times are investigated.  In the next section, the various scattering mechanisms used in MC simulations are described.  In the third section, MC simulation results are presented, followed by discussions of the results.  In the last section, we draw conclusions and discuss possible future work.

\section{Scattering mechanisms in graphene}

At finite temperature, the {\em intrinsic} scattering mechanisms that affect carrier transport are due to electron-phonon interactions.  The phonon modes that need to be accounted for are acoustic and optical.  The scattering rates are calculated using Fermi's Golden rule \cite{transportbook_Lundstrom}.  The acoustic phonon scattering rate from state $|k\rangle$ to state $|k'\rangle$ is given by
\begin{equation}
S_{ac}(k,k') = \frac{\pi D_{ac}^2 k_{B}T (1+\cos{\theta_{kk'}})}{ 2 \hbar \sigma_{m} v_{p}^2} \delta(\mathcal{E}_{k}-\mathcal{E}_{k'}),
\end{equation}
where $D_{ac} \sim 16$ eV is the deformation potential of acoustic phonon scattering, $k_{B}$ is the Boltzmann constant, $T$ is the temperature, $\mathcal{E}_{k} = \hbar v_{F}|k|$ is the kinetic energy of the carrier and $k$ is the wavevector, $\hbar$ is the reduced Planck constant, $v_{F} = 10^{8}$ cm/s is the Fermi velocity in graphene, $\sigma_{m} \sim 7.6\times 10^{-8}$ g/cm$^2$ is the 2D mass density of graphene, and $v_{p}=20$ km/s is the acoustic phonon (sound) velocity in graphene.  $\theta_{kk'}$ is the angle of electron velocity between initial state $|k\rangle$ and final state $|k'\rangle$.  Acoustic phonon scattering in graphene is not isotropic, since the acoustic phonon perturbation is long-range, and the wave functions of carriers in graphene have the `pseudospin' symmetry.  The angle dependent factor for acoustic phonon scattering is $1+\cos{\theta}_{kk'}$, which prevents backscattering.  The acoustic phonon scattering rate for electron in state $|k\rangle$ is calculated by summing over all possible finals in $\bf{k}$ space and it is given by
\begin{eqnarray}
\frac{1}{\tau_{ac} (k) } = \frac{D_{ac}^2 k_{B}T }{ 2 \hbar^3 v_{F}^2 \sigma_{m} v_{p}^2} \times \mathcal{E}_{k},
\end{eqnarray}
where acoustic phonon scattering is treated as quasi-elastic, and both emission and absorption are considered.

The optical phonon scattering rate from state $|k\rangle$ to state $|k'\rangle$ is given by
\begin{eqnarray}
S_{op} (k,k') = \frac{\pi D_{o}^2 }{\sigma_{m}\omega_{o}}[n_{op}\delta(\mathcal{E}_{k}-\mathcal{E}_{k'}-\hbar\omega_{o}) \\ \nonumber
+(n_{op}+1)\delta(\mathcal{E}_{k}-\mathcal{E}_{k'}+\hbar\omega_{o})],
\end{eqnarray}
where $D_{o}$ is the mode-specific optical deformation potential of graphene, $\omega_{o}$ is the optical phonon frequency and $n_{op}=1/(e^{\hbar\omega_{o}/k_{B}T}-1)$ is the optical phonon number.  Similarly, the optical phonon scattering rate for state $|k\rangle$ is the summation of all scatterings to final states, and it is given by
\begin{equation}
\frac{1}{\tau_{op} (k)} = \frac{D_{o}^2 (n_{op}+\frac{1}{2}\mp \frac{1}{2})}{2\hbar^2v_{F}^2\sigma_{m}\omega_{o}} \times (\mathcal{E}_{k} \pm \hbar\omega_{o}),
\end{equation}
where $\pm$ is for phonon absorption and emission respectively.  In graphene, the zone edge (ZE) transverse optical (TO) mode has the strongest coupling with electrons \cite{PrB09_RG_TOphonon} and the energy of this mode is around $\hbar \omega_{o} = 160$ meV.  The `optical' deformation potential of the mode is $D_{o}=25.6$ eV/\AA\ \cite{nanoLett_CNTtransport_Cornell}.  The scattering of zone-edge mode is non-isotropic and has the angle dependence factor $1-\cos{\theta}_{kk'}$ \cite{jpsj2006_opticalphnon_CNT}.  Forward scattering ($\theta=0$) is prohibited and back scattering ($\theta=\pi$) is preferred due to this factor; the situation is opposite of what happens in acoustic phonon scattering.  In our simulation, we also include the zone center (ZC) optical phonon.  The ZC optical phonon has a higher energy, $\hbar \omega_{o} = 196$ meV.  The deformation potential for the ZC phonon is
$D_{o}=14.1$ eV/\AA  \cite{prl07_fieldtuningphonon_Columbia}.  The zone center optical phonon scattering is isotropic \cite{jpsj2006_opticalphnon_CNT}.

\begin{figure}
\begin{center}
\leavevmode
\epsfxsize=2.8in
\epsffile{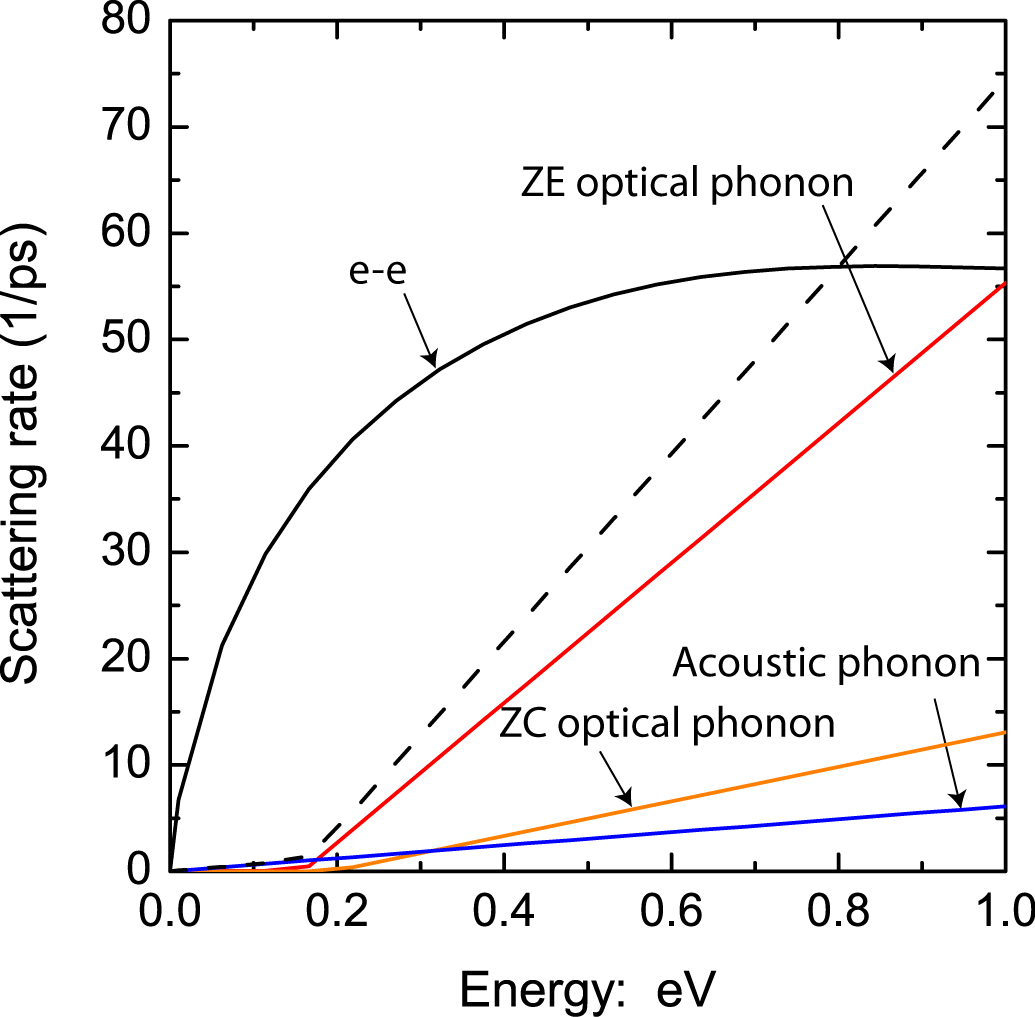}
\caption{Scattering rates versus energy in 2D graphene: T=300 K, $n_{e}=7.7\times 10^{12}$ cm$^{-2}$. The dashed line is the sum scattering of acoustic phonon and optical phonons.}
\label{scatteringrates}
\end{center}
\end{figure}

The effect of e-e scattering in graphene with linear $\mathcal{E} \sim k$ dispersion is different from traditional semiconductors that have parabolic band structure ($\mathcal{E} \sim k^{2}$) \cite{apl10_eescattering_graphene}.  Energy and momentum conservation requirements modify the carrier distribution and center of mass motion in a way that the low-field transport properties (such as carrier mobility) is strongly damped, especially at low carrier concentrations. The e-e scattering rate of electron at state $|k\rangle$ with another electron at $|k_{0}\rangle$ is given by
\begin{eqnarray}
S_{ee}(k,k_{0})=\frac{2\pi}{\hbar}(|V_{q}|^2+|V_{q'}|^2+|V_{q}-V_{q'}|^2)\\ \nonumber
\delta_{k+k_{0},k'+k_{0}'}\delta(\mathcal{E}_{k}+\mathcal{E}_{k_{0}}-\mathcal{E}_{k'}-\mathcal{E}_{k_{0}'}),
\label{eescatteringrate}
\end{eqnarray}
where $k'$ and $k_{0}'$ are the finals states.  The Coulomb interaction matrix $V_{q}$ is given by
\begin{equation}
V_{q} = \frac{e^2}{2\epsilon(q)q} \frac{(1+\cos{\theta_{kk'}})}{2}\frac{(1+\cos{\theta_{k_{0}k_{0}'}})}{2},
\end{equation}
where wave vector $q=|k-k'|$ and $\epsilon(q)$ is the dielectric function of 2D graphene.  $\theta_{k_{1}k_{1}'}$ and $\theta_{k_{2}k_{2}'}$ are the scattering angles of the two electrons, which has to be taken into account since Coulomb scattering is also long range interaction.  Similarly, wave vector $q'=|k-k_{0}'|$ is for the scattering process of exchange of two electrons.  The two delta function shows the conservation of momentum and energy in the scattering.  The electron scattering rate is calculated by summation of all $k_{0}$ state and also all the final states ($k'$,$k'_{0}$), and it is given by
\begin{eqnarray}
\frac{1}{\tau_{ee}(k)}=\frac{2\pi}{\hbar}\sum_{k_{0}} f(k_{0})\sum_{(k',k_{0}')}(|V_{q}|^2+|V_{q'}|^2+|V_{q}-V_{q'}|^2)\\ \nonumber
\delta_{k+k_{0},k'+k_{0}'}\delta(\mathcal{E}_{k}+\mathcal{E}_{k_{0}}-\mathcal{E}_{k'}-\mathcal{E}_{k_{0}'}),
\label{eetotalrate}
\end{eqnarray}
where the occupation of state $k_{0}$ is the Fermi-Dirac distribution and the final states are assumed to be empty.

The e-e scattering rate is incorporated into the MC simulation in this work following the procedure of ref.\cite{apl10_eescattering_graphene} with one difference: the screening of Coulombic forces is treated as temperature-dependent \cite{prB09_Tdependentscreening_Maryland}.

The scattering rates at 300 K are plotted in Fig.\ref{scatteringrates}.  In order to capture high-field effects, the scattering rate from 0 to 1 eV are calculated for all the scattering mechanisms in intrinsic graphene.  The intrinsic optical phonon scattering rate increases linearly with energy at states higher than $\hbar \omega_{O}$.  Acoustic phonon scattering rate is one order of magnitude smaller than the optical phonon scattering at high energies.  Electrons are accelerated by electric field and populate high-energy states.  Therefore, optical phonon scattering is the dominant energy relaxation channel for hot electrons rather than acoustic phonon.  Since the ZE optical phonon has lower energy and higher scattering rate, it is the dominant optical mode.

The e-e scattering rate at 300 K is higher than optical phonon scattering.  This scattering thermalizes the carrier distribution, driving it towards a Fermi Dirac distribution.  The net momentum does not change during e-e collisions, but the average velocity of electrons changes due to the linear band structure of 2D graphene.  As a consequence, e-e scattering has an impact on the current.  The e-e scattering rate increases with carrier temperature so that it become more important under high field.

From Fig.\ref{scatteringrates}, the phonon scattering rate at high energies exceeds 10 ps$^{-1}$.  Optical phonons emitted by electrons have to relax to other phonon modes by anharmonic interaction before the heat propagates into the substrate or contacts.  The characteristic lifetime of optical phonon decay into acoustic phonon modes is $\sim$1 to 5ps in carbon-based $sp^{2}$ crystals \cite{prb06_hotphononinCNT}.  Since the optical phonon lifetime is much longer than the phonon generation time ($\sim$ 0.1 ps), the optical phonon population in graphene is out of equilibrium.  The fast generation rate and slow decay rate of optical phonons result in a hot phonon effect in graphene.

Before presenting the numerical results of the ensemble MC calculation, a heuristic picture of the high field transport properties in graphene is developed in the next section.  A simple model considering the effect of carrier degeneracy provides analytical estimation of the saturation currents for degenerate electron systems.


\section{High field transport model in graphene}

\begin{figure}
\begin{center}
\leavevmode
\epsfxsize=3.6in
\epsffile{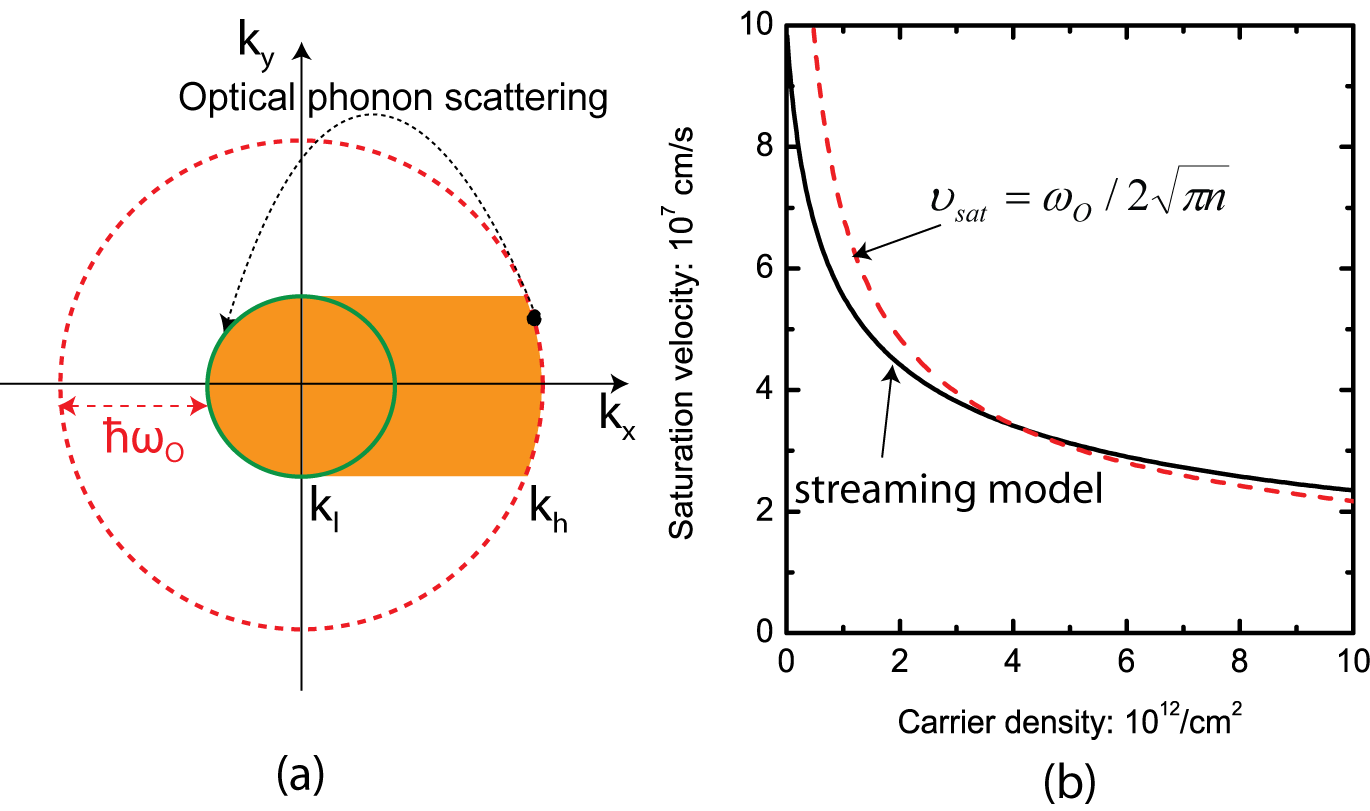}
\caption{Current saturation model in 2D graphene: (a) The distribution function of streaming model at steady state.  f($\bf{k}$)=1 at colored region, otherwise, f($\bf{k}$)=0. (b) The saturation velocity versus carrier concentration.}
\label{mymodel}
\end{center}
\end{figure}

The electron distribution function in the $\bf{k}$- space in the steady state at high electric fields can be approximated since the optical phonon emission is much faster than competing energy relaxation mechanisms.  The steady state electron distribution function $f(\bf{k})$ under this assumption is shown in Fig.\ref{mymodel}(a).  The electrons are distributed between two energy contour circles, the high energy circle $\mathcal{E}_{h}=\hbar v_{F}k_{h}$ and the low energy circle $\mathcal{E}_{l}=\hbar v_{F}k_{l}$.  All electrons move toward $k_{x}$ direction under the applied electric field.  Once electrons approach the high energy circle on the right, they are scattered back to low energy circle by emitting optical phonon instantaneously.  Therefore, the energy difference between the two circles is the optical phonon energy, $\mathcal{E}_{h}-\mathcal{E}_{l}=\hbar \omega_{o}$, as indicated in the figure.  The state occupation inside the low energy circle is full so that optical phonon emission by electrons with energy smaller than $\mathcal{E}_{h}$ are prohibited by the Pauli exclusion principle.  As optical phonon emission continues and scatters the electrons from high energy states back to the low energy states, distribution function $f(\bf{k})$ is squeezed and elongated along the direction of the electric field.  The steady distribution function is called `streaming' distribution.  This streaming model captures the effect of carrier degeneracy, which has not been considered in high field-transport in traditional non-degenerate semiconductors.

The carrier concentration is proportional to the area of the distribution function $\mathcal{S}$ in $\bf{k}$ space, $n=g_{s}g_{v}\mathcal{S}/(2\pi)^2$, where $g_{s}$ and $g_{v}$ are spin and valley degeneracies in graphene.  The area of the distribution function is given by
\begin{equation}
\mathcal{S}=\frac{\pi k_{l}^2}{2}+k_{l}\sqrt{k_{h}^2-k_{l}^2}+k_{h}\tan^{-1}(\frac{k_{l}}{\sqrt{k_{h}^2-k_{l}^2}}),
\label{area}
\end{equation}
from which the dimensions of the distribution function ($k_{l}$,$k_{h}$) can be calculated at any carrier concentration. The current flowing in the direction of the electric field corresponding to the distribution function in Fig.\ref{mymodel}(a) is then found by summing the $x-$directed group velocities of all states inside the shaded region in the $\bf{k}$ space; the value is found to be
\begin{equation}
J_{sat}=\frac{2e\omega_{o}}{\pi^2\hbar v_{F}}\mathcal{E}_{l},
\label{Jsatofstreamingmodel}
\end{equation}
where $\mathcal{E}_{l}$ is the electron energy at low energy circle.  This streaming model shows that the saturation current is proportional to the optical phonon frequency and radius of low energy circle.  In Fig.\ref{mymodel}(b), the ensemble saturation velocity $v_{sat} = J_{sat} / en$ is plotted against the electron density.  The saturation velocity varies with carrier concentration and is not a constant.  It approaches the Fermi velocity in graphene $v_{F} = 10^{8}$ cm/s as $n\rightarrow 0$, and falls to $\sim 2 \times 10^{7}$ cm/s at $n\sim 10^{13}$/cm$^{2}$.  The dotted line shows the closely related square root relation $v_{sat}=\omega_{o}/2\sqrt{n\pi}$ proposed in ref.\cite{DRC08DJ}.

Traditionally, the saturation velocity in non-degenerate semiconductors has been considered a constant, e.g. $v_{sat}=10^{7}$ cm/s in Si at room temperature \cite{SSE_77_SiVsat}.  The reason for the carrier-density dependent saturation velocity in graphene is the degeneracy of carriers and the Pauli exclusion principle.  Close to inverse square-root dependence of the effective ensemble saturation velocity on the carrier density has been measured in graphene on SiO$_{2}$ substrate \cite{08natnano, DRC2010_Pop}.  The model presented here captures the essence of the high-field transport and is in agreement the experimental results, indicating that the carrier degeneracy plays a major role on high field transport in graphene.  However, the model here can stand on a firmer footing by using a numerical Monte-Carlo approach to find high field transport properties, which is the topic of the next section.


\section{Monte Carlo simulation: Implementation and results}

In the MC implementation of the high-field transport problem, we have considered electron energies ranging from 0 to 10 $\hbar \omega_{o}$ in the graphene conduction band.  The $\bf{k}-$space is divided into around $10^{5}$ cells.  In simulation, `superelectrons' move in $\bf{k}$ space according to Newton's law ${\bf F} = \hbar d{\bf k}/dt$ \cite{transportbook_Lundstrom}.  The number of `superelectrons' varies from 1000 to be more than $10^{4}$, depending on the carrier density.  The current is calculated by summation of the group velocities over all the `superelectrons'.

The screening of the e-e Coulombic interaction is dependent on the carrier temperature.  In the MC simulation, the electron temperature $T_{e}$ is updated every time slot (\~0.5 fs).  The temperature was calculated from the thermal energy of electron gas $\mathcal{E}_{th}=\hbar v_{F}\sum_{i}(k_{i}-\bar{k})$, where $k_{i}$ is wavevector of the i'th electron, and the average wavevector of all electrons is $\bar{k}$.  The electron temperature was calculated by assuming a Fermi-Dirac distribution, whereby the thermal energy of the electron gas at temperature $T_{e}$ is given by
\begin{equation}
\mathcal{E}_{th} = \frac{2A}{\pi}\frac{(k_{B}T_{e})^3}{(\hbar v_{F})^2}\int_{0}^{\infty}\frac{\mu^2d\mu}{1+e^{\mu-\eta}},
\end{equation}
where $A$ is the area of graphene sheet and $\eta=\mathcal{E}_{f}/k_{B}T_{e}$.  In simulation, we assume the carrier density in graphene conduction band is fixed by ignoring generation processes.  As a result, Fermi level changes with the temperature, and both of them can be calculated once we know the thermal energy and carrier concentration.

In order to capture the effect of hot phonons, the phonon ${\bf q}$-space is discretized in the same fashion as the electron ${\bf k}$-space, and the hot-phonon effect is treated as described in \cite{PrB05_AlGaN_hotphonon}.  The optical phonon distribution function is updated based on the phonon wavevectors during the emission and absorption events.  Optical phonon decay into other phonon modes is governed by the rate equation of phonon occupation function $n_{q}(t)$ -
\begin{equation}
n_{q}(t+\delta t)=n_{q}(t)-\frac{n_{q}(t)-n_{q0}}{\tau_{ph}}\delta t,
\end{equation}
where $\delta t$ is the time slot in MC simulation, $n_{q0}$ is the equilibrium Bose-Einstein distribution, and $\tau_{ph}$ is the phonon lifetime.  The time slot $\delta t$ is around 0.5 fs, corresponding to a scattering rate 2$\times 10^{3}$ ps$^{-1}$, which is higher than the summation of all scattering rates for electrons.  This time slot was updated according to the phonon population in simulation, since the maximum scattering rate varies with the phonon population.

The degeneracy of carriers is explicitly dealt with by the `rejection technique' \cite{TED85_MC_degeneracy}.  The simulation is continued till a steady state distribution function is obtained.  This procedure yields the distribution functions of electrons $f(k_{x}, k_{y}, t)$ and phonons $n(q_{x}, q_{y}, t)$ at every time step, which is then used to find ensemble velocities, currents, as well as transient effects such as velocity overshoot as a function of the carrier density and the applied electric field.
\begin{figure}
\begin{center}
\leavevmode \epsfxsize=3.6in \epsffile{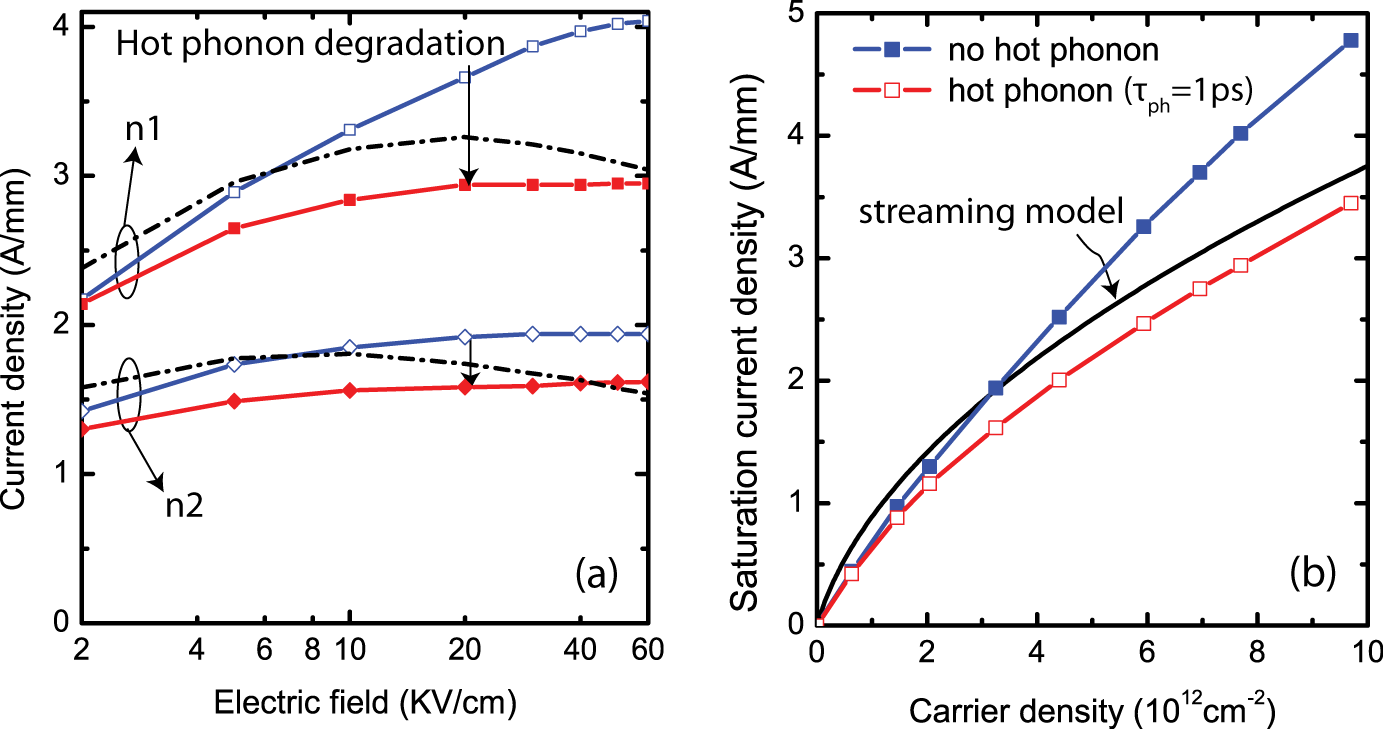}
\caption{(a) The current density versus electric field at two carrier concentration n1=$7.5\times 10^{12}$ cm$^{-2}$ and n2=$3.0\times 10^{12}$ cm$^{-2}$.  Open symbols: no hot phonon; Solid symbols: hot phonon ($\tau_{ph} = 1$ ps).  Dot-dash lines consider hot phonon effect but not e-e scatterings.  (b) Saturation current versus carrier density. Solid black curve: the current saturation results of streaming model;  Solid square: intrinsic graphene without hot phonon effect, $\mathcal{F}=50$ kV/cm; Open square: with hot phonon effect ($\tau_{ph}$=1 ps), $\mathcal{F}=50$ kV/cm.  All simulations are at 300 K.}
\label{MCresults}
\end{center}
\end{figure}

The steady-state currents calculated by the MC simulation for various strengths of electric fields are shown in figure \ref{MCresults}(a) for two carrier densities.  The curves of different carrier concentrations show similar behavior.  The currents increase with the field and saturate at a field $\mathcal{F} \sim 50$ kV/cm.  The degradation of the saturation current due to the hot-phonon effect is clearly seen by comparing the open and solid symbol lines, which is indicated by the arrows.  Hot phonon effect is more significant in degradation of the saturation current at higher carrier concentrations.  In order to highlight the effect of e-e scattering at high field, the transport without e-e scattering is also simulated.  The dot-dash lines are the currents including hot phonon effect but without e-e scatterings.  Without e-e scattering, the currents drop at high fields, which is the indication of negative differential resistance (NDR).  NDR has been proposed and observed in simulations for carbon naotubes (CNTs) and graphene \cite{prl05_NDRsemiconductingCNT_IBM,apl06_NDRmatellicCNT_UIUC,jpcm09_NDRgraphene_Ferry}, due to the linear dispersion of band structure.  In graphene, the electrons at high energy state move at the same speed with the low energy state, which is the constant Fermi velocity in graphene.  At high field, more electrons move into high energy states.  But the increasing of velocity is not as rapidly as conventional parabolic materials.  As more and more electrons occupy high energy states the backscattering is increased, which degrades the average velocity of electron gas and leads to NDR.  However, this effect disappears (or weakened) if e-e scattering is considered.  The e-e scatterings exchange electrons' momentum and energy, diving electrons into Fermi-Dirac distribution.  As a result, high energy electrons exchange their momentum and energy with low energy electrons before substantial backscatterings happen.  With e-e scatterings, the hot electrons generation and backscattering are weakened and the ensemble velocity of carriers increases with the applied electric field.  The current-field curves from the MC simulation therefore show that a) the hot phonon effect acts as a strong current limiting mechanism at high carrier densities and b) e-e scattering gets rid of NDR effect in graphene.

In Fig.\ref{MCresults}(b), the saturation current is plotted against the carrier density.  The solid black curve is the analytical result from streaming model (equation \ref{Jsatofstreamingmodel}) for comparison.  The solid squares are the intrinsic saturation currents without the hot phonon effect.  The saturation current is slightly lower than the analytical result due to hot phonon degradation.  Saturation current density of $\sim$3 A/mm has been measured in epitaxial graphene on SiC substrate ($n \approx 10^{13}$ cm$^{-2}$) \cite{EDL09grfet_HRL}, which is consistent with the MC result.  The current-carrier concentration curves are not linear, indicating the ensemble saturation velocity is not constant, but decreases with increasing carrier density.

\begin{figure}
\begin{center}
\leavevmode \epsfxsize=3.6in \epsffile{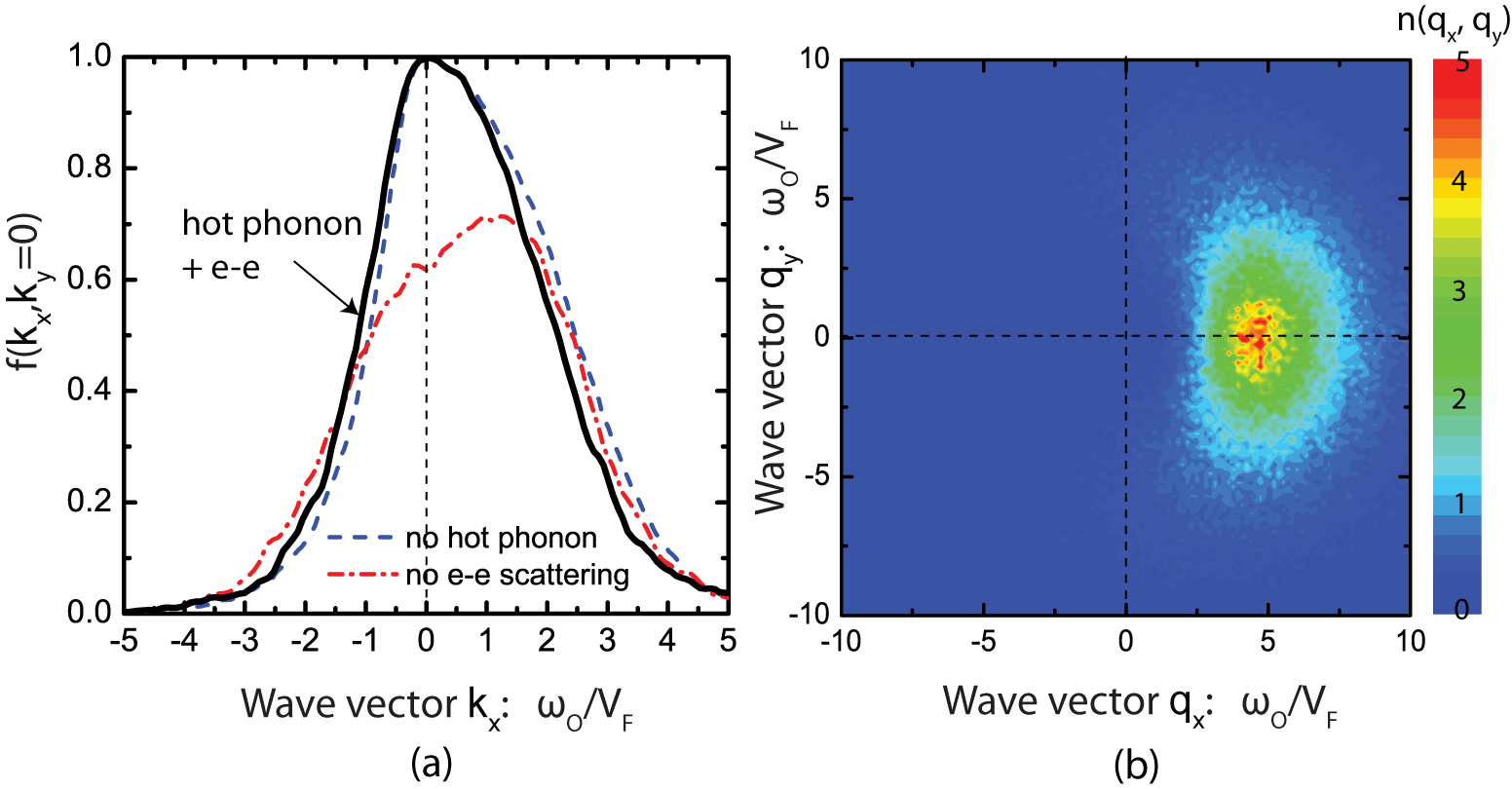}
\caption{(a) The electron distribution function along $k_{x}$ with $k_{y}$=0. Solid line considers hot phonon and e-e scattering.  (Dash line: turning off hot phonon effect and dash-dot line: turning off e-e scattering in simulation) (b) The phonon number distribution function in $\bf{q}$ space.  Carrier density: n=7.5$\times 10^{12}$ cm$^{-2}$, electric field $\mathcal{F}=50$ kV/cm, phonon lifetime $\tau_{ph}$=1 ps, environment temperature $T=$300 K.}
\label{distributionfunctons}
\end{center}
\end{figure}

The hot phonon effect is better illustrated by the distribution functions of electrons in ${\bf k}$-space, and phonons in ${\bf q}$-space.  The electron distribution functions are plotted in Fig.\ref{distributionfunctons}(a) along the $k_{x}$ axis which is along the direction of the electric field.  The solid line is the electron distribution at steady state, considering hot phonon and e-e scattering.  The peak of the distribution function is close to 1, showing the high degeneracy of carriers in graphene.  For comparison, we turned off the e-e scattering or hot phonon effect in the simulation.  The distribution function without e-e scattering has a lower peak and is broader, shown by the dot-dash curve.  Compared to the solid distribution curve, the distribution without e-e scattering has more electrons at high energy states, but also more backscattered electrons.  This highlights the role of e-e scattering in maintaining the shape of carrier distribution close to Fermi-Dirac like (the solid line) and thus preventing the NDR effect.  Similarly, the effect of hot phonons becomes clear when we compare the distribution functions shown by the solid and dashed curve.  The carrier distribution with hot phonons is shifted less along the field direction, so the saturation current is lowered by the hot-phonon effect.

The effect of hot phonons on the transport can also be seen from the phonon distribution in ${\bf q}$-space.  Fig.\ref{distributionfunctons}(b) shows the 2D phonon distribution functions $n(q_{x}, q_{y})$.  The hot phonon distribution has a peak at positive $q_{x}\approx 5\omega_{O}/v_{F}$, which corresponds to the fast optical phonon emission by hot electrons.  The phonon occupation number peak exceeds $n_{ph} \sim 5$ which corresponds to a phonon temperature of $T_{op} \sim 10^{4}$ K.

\begin{figure}
\begin{center}
\leavevmode \epsfxsize=3.6in \epsffile{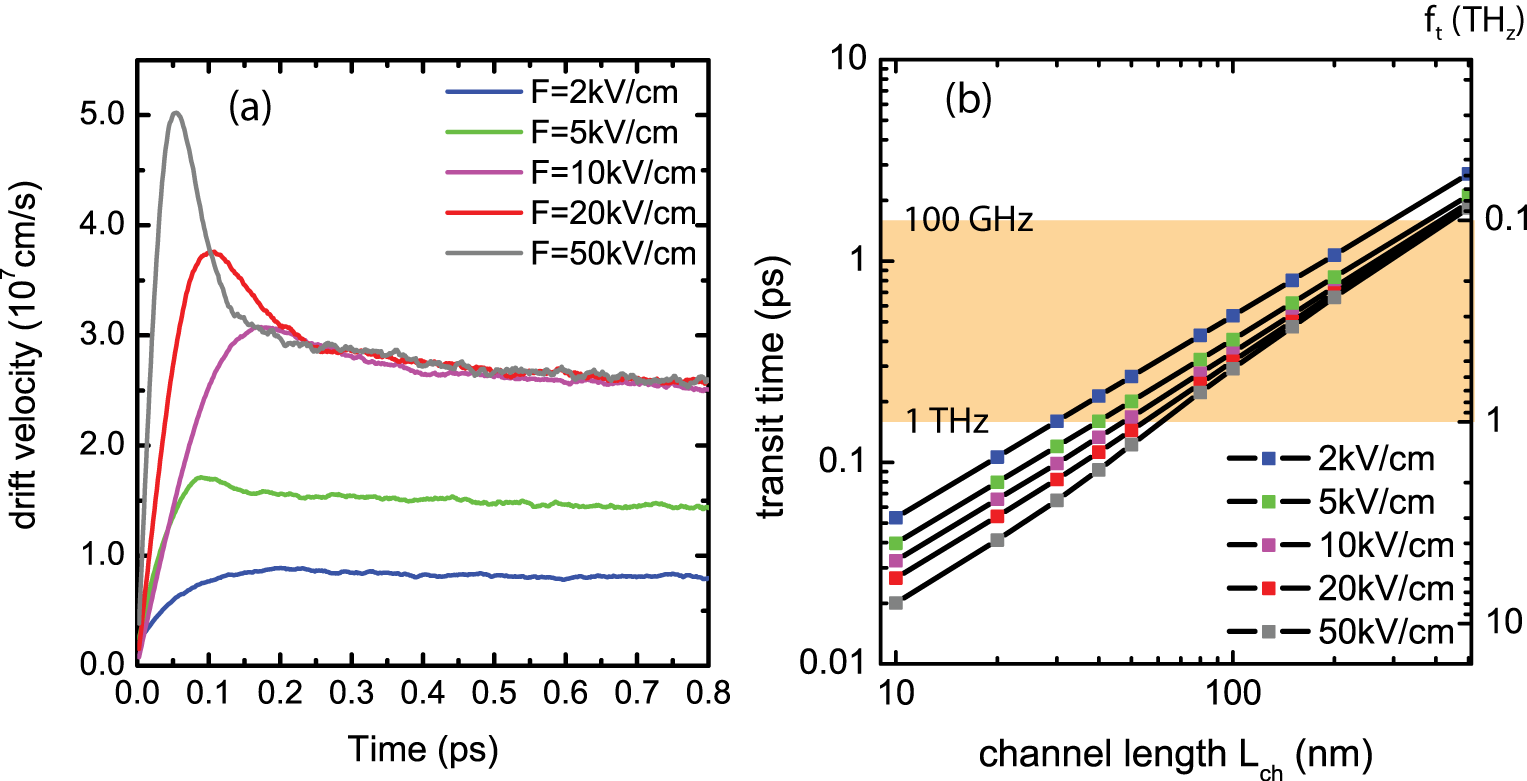}
\caption{(a) Electron drift velocity evolution with time under field(b) Carrier transit time versus channel length. Optical phonon lifetime $\tau_{ph}$=1 ps, environment temperature $T$=300 K, carrier density $n=7.5\times 10^{12}$ cm$^{-2}$.}
\label{VdvsTime}
\end{center}
\end{figure}

Besides the steady-state saturation currents, the MC simulation procedure also allows us to study transient phenomena in high-field carrier transport.  Prior to approaching steady state, carriers moving short distances in traditional semiconductors are known to undergo transient velocity overshoot, attaining higher velocities than what can be attained in the steady state.  The hot electrons accelerated by high field keep moving in $\bf{k}$ space without significant phonon emission during timescales shorter than the mean phonon emission time.  As a result, a velocity peak is achieved in a short period.  In high speed electronic devices, the channel length is scaled to tens of nanometers.  The carriers travel from source to drain contacts within ps timescales, and velocity overshoot can occur.  This transient velocity overshoot can be utilized to lower the transit time of carriers in graphene devices, thus resulting in an enhanced speed.  Fig.\ref{VdvsTime}(a) shows the ensemble transient velocity response to electric fields.  Velocity overshoot occurs when the electric field is higher than 5 kV/cm.  The higher the field, the higher is the overshoot velocity peak; it approaches $\sim 5 \times 10^{7}$ cm/s at a field of 50 kV/cm for the chosen carrier density ($n\sim 7.5 \times 10^{12}$/cm$^{2}$).  After reaching the peak overshoot velocity, the ensemble drift velocity decreases gradually by dissipating excess energy into the graphene crystal in the form of optical phonon vibrations.  The long tail of decreasing velocity is due to the hot phonon effect - optical phonons are slow in decaying into acoustic modes which heat up the lattice.

Velocity overshoot occurs within a time window of $t_{os} \sim 0.2$ ps.  If the channel length ($L$) is short enough ($L < L_{os} = \int_{0}^{t_{os}} v_{d} (t) dt $), the hot carrier energy is dissipated at the drain contact, and the hot phonon effect can be prevented in a short channel device.  Thus, the distance $L_{os} = \int_{0}^{t_{os}} v_{d} (t) dt $ that carriers travel in the velocity overshoot period is crucial for the design of high-speed devices.  The transit times $t_{tr}$ given by $L = \int_{0}^{t_{tr}} v_{d}(t) dt$ is calculated from the result in Fig\ref{VdvsTime}(a) and shown for various lengths in \ref{VdvsTime}(b).  For short channel lengths, the transit time is sensitive to the magnitude of the electric field, since velocity overshoot increases with electric field.  For long channel lengths, a major portion of the transit time is spent traveling at the steady state drift velocity, and the overshoot velocity makes a minor contribution.  In the MC simulation, the transit time is not found to be particularly sensitive to the carrier density, which bodes well for devices requiring high currents and high speeds.  In a $L = 80$ nm long channel device, the transit time is found to be $t_{tr} \sim 0.2$ ps, which is exactly the time window for velocity overshoot.  Thus, graphene channels $L \leq 80$ nm can fully take advantage of velocity overshoot for enhancing their speed.

The peak overshoot velocity, and overshoot time window ($t_{os}$) is determined by strength of electron-optical phonon interaction, the optical phonon energy, as well as the low-field mobility. The optical phonon energy and carrier-phonon interaction are intrinsic properties of material, but the peak overshoot velocity peak can be improved by increasing the cleanliness of the substrate or by using gate dielectrics with less impurities.


\section{Discussion and conclusion}

We point out some limitations of this work, and suggest topics that need investigation in the future.  In this work we have ignored the band to band processes, e.g. inter-band (Zener) tunneling and e-e scattering induced carrier generation.  The carrier concentration does not change at high electric field under the assumption. However, this is not the case in graphene if the Fermi level is close to the Dirac point.  Since the band gap is zero, tunneling and e-e scatterings generate electrons and holes under high fields at very low carrier concentrations.  In experimental measurements, the current increases and never saturates when the Fermi level is close to the Dirac point in intrinsic graphene ($\mathcal{E}_{f}\sim0$) \cite{prl_highfield_spain}.  Field-induced carrier generation is also the reason of the low on/off ratio ($\sim 10$) in graphene FETs.

In experiments, intrinsic properties of graphene is difficult to achieve.  SO phonon scattering limits the high field transport for graphene in close proximity to substrates with low energy phonon modes, like SiO$_{2}$.  Suspended graphene is free from environmental scattering effects, but the carrier density is not easy to modulate.  One possible way to measure the intrinsic performance is to use SiC substrate graphene which has weak SO phonon coupling.

We assumed the lattice temperature to be at 300 K in our simulations, although the optical phonon temperature can be much higher than environment temperature 300 K.  This assumption could overestimate the saturation current, since the lattice temperature in graphene increases due to the limited thermal conductance of the substrate and contacts.  The hot phonon effect is dominant, so to a certain extent ignoring the rise of lattice temperature is justified.

The phonon lifetime has been measured for graphene on SiC substrate and it was found to be around 2.5 ps \cite{condmat09_OpLifetimeMeasurement_Farhan}.  The hot phonon effect can be diminished by introducing isotopic disorder into graphene \cite{apl08_hotphonon_GaN}.  Isotopic disorder introduces phonon modes that are localized in real space, and therefore spread in $\bf{q}$-space, and more phonon modes are thus involved in the cooling of carriers.  The saturation currents can be improved by this disorder engineered lifetime of optical phonons, with direct consequences of higher saturation currents.

In conclusion, we investigated the high field transport in intrinsic monolayer graphene.  MC simulation results show that both hot-phonon effect and e-e scattering have strong impacts on the saturation current.  Transient velocity overshoot was studied and the short carrier transit time indicates that graphene is well-suited for high speed electronics.

\section{Acknowledgements}
The authors acknowledge Kristof Tahy and Prof. Eric Pop's group for discussions and experimental data on high field transport in graphene, the NSF CAREER and ECCS awards, and the SRC NRI MIND center for financial support.


\end{document}